\documentstyle[12pt]{article}
\begin{document}
\begin{titlepage}
\vspace*{2.7cm}
\begin{center}
{\Large The Decays of Glueballs to two light mesons\\}
\vspace*{0.4cm}
{{J.Y. Cui $^a$ $^ b$  H.Y. Jin $^a$ and J.M. Wu $^a$}\\
  {$^a$ \small Institute of High Energy Physics}\\
  {\small Beijing, 100039, P.R. China}\\
  {$^b$ \small Department of Physics, Henan Normal University}\\
  {\small Xinxiang, 453002, P.R. China}}
\end{center}
\date{}
\begin{center}
\begin{minipage}{120mm}
\vspace*{1.5cm}
\begin{center}{\bf Abstract}\end{center}
{ We solve the B.S. equation of glueballs under instantaneous approximation.
With the B.S
wave function obtained we calculate the decay width of glueballs to two
pseudoscalar  mesons, $\Gamma(\pi\pi)$, $\Gamma(kk)$ and $\Gamma(\eta\eta)$. 
$\Gamma(\eta\eta^\prime)$ from QCD anomaly is also estimated. \\
{\bf PACS: 14.40.C, 14.40.G, 11.10.St, 13.25 \\
Keywords: glueball, B.S. equation, QCD anomaly, decay}}
\end{minipage}
\end{center}
\vskip 1in
\end{titlepage}
\newpage
1.{\it Introduction}${ ~~} $ The existence of the glueball states is the
prediction of QCD, whose discovery would be a direct confirmation of the
non-ablian character of QCD. A lot of investigation both 
in theory [1--5] and in 
experiment [6--8] have been made in the glueball searches. At present 
there are 
several experimental glueball candidates, such as $f_0(1500)$, $f_J(1700)$ and
$\xi(2230)$, but non is identified as glueball definitely. It's  difficult
to identify a glueball because so far we haven't understood the glueball
clearly in theory: we don't know clearly its mass, its production rates and 
its decay width. Therefore, more investigations are necessary.

In this paper, we focus on $0^{++}$ glueball's decays. Up to
now, $f_0(1500)$ is considered as the lowest-lying $0^{++}$ glueball 
candidate. One of the most important reason is that the branching
 ratio of decay 
$f_0(1500)\rightarrow \eta\eta'$ is surprisingly large, i.e
$\Gamma(\eta\eta')/\Gamma(\eta\eta')\sim 1$. Since not only $\eta\eta'$ 
 final state is almost on the kinematic threshold but 
also it is suppressed by SU(3) flavor symmetry, 
the naive quark model gives 
$\Gamma(\eta\eta')/\Gamma(\eta\eta')\approx 0$. Therefore $\eta'$ must be 
produced by another mechanism. One of the possible mechanism is the 
famous QCD anomaly. However, it is well known  that QCD anomaly is large
$N_c$ suppressed, in order to produce enough rate of $\eta'$ production, 
$f_0(1500)$ should have a high gluon density, i.e., it is a glueball.
On this assumption , the further theoretical 
calculation is still necessary to check the rate
$\Gamma(\eta\eta')/\Gamma(\eta\eta)$ and
$\Gamma(\eta\eta')/\Gamma(\pi\pi)$. We notice there have already been 
some works on this topic from lattice QCD approach, however, 
an model-dependent calculation is still useful and may give some helpful 
hints.       

\par  In an early paper \cite{s5}, we studied the glueball spectrum in the
framework of
B.S. equation, and the results are in good agreement with that of Lattice
calculation. This encourages us to do more investigation. By solving the B.S.
equation about gluon-gluon bound state, we can obtain not only the  glueball
spectrum but also the glueball B.S. wave function. In this paper we try to use
these wave functions to estimate the decay width of glueballs to light
pseudoscalar mesons in the framework of perturbative QCD.

\par This paper is arranged as follows. In the next section we briefly
review the construction of the
B.S. equation for gluons bound state.  In section three, first we calculate the
glueball width to two pions. Then we give the estimations of width to two
K's and two $\eta$'s.
In section four, we give the estimation of $\Gamma(\eta\eta^\prime)$ from QCD
anomaly. In the last section a discussion is presented.
\par
2.{\it The Glueball B.S. Equation }${ ~~ }$ The details of the
construction of the glueball B.S. equation has been given in ref.\cite{s5}. Here
we only present a brief review. 
\par
  Let $A^a _\mu (x_1)$ and $A^b _\nu (x_2)$ be the gluon fields at
  points $x_1$ and $x_2$,
$|G \rangle $ the bound state of two gluons with mass $M_G$ and momentum $P_\mu$.
Then the B.S. wave function for a bound state is defined as
\begin{equation}
\chi^{a b}_{\mu \nu}(P,x_1,x_2)=\langle 0 |T(A^a _\mu (x_1) A^b
_\nu (x_2))| G \rangle,
\end{equation}
where $\mu ,  \nu $
are Lorentz indices, and $a, b$ color indices. The glueballs are color singlet 
states, so we have
\begin{equation}
\chi^{a b}_{\mu \nu}(P,x_1,x_2)=\delta^{ab}\chi_{\mu \nu}(P,x_1,x_2).
\end{equation}
With a standard method we obtain the B.S. equation for a color singlet
glueball state,
\begin{equation}
\chi_{\mu \nu}(P,q)=\Delta_{\mu \alpha}(p_1)\Delta_{\nu
\beta}(p_2)\int\frac{d^4 k}{(2\pi)^4}G^{\alpha \beta \rho
\sigma}(P,q,k)\chi_{\rho \sigma}(P,k),
\end{equation}
where $\chi_{\mu \nu}(P,q)$ is the Fourier transformation of
$\chi^{a b}_{\mu \nu}(P,x_1,x_2)$.

 The tensor kernel $G^{\alpha \beta \rho \sigma}(P,q,k)$, in which 
the color indices are suppressed, 
is defined as the sum of all two-particle irreducible graphs,
and $\Delta_{\mu \alpha}(p_i)$ is the full propagator of the constituent gluons
with momentum $p_i$.
\begin{equation}
\label{e2a}
 P=p_1+p_2,{~~~~}2q=p_1-p_2,
\end{equation}
\par
Under instantaneous approximation, we ignore the $q_0$ and $k_0$ dependence
of the kernel $G^{\alpha \beta \rho \sigma}(P,q,k)$. Further we replace
$\Delta_{\mu \alpha}(p_i)$ with a free propagator. Then in Coulomb gauge we
obtain a three dimensional equation
$$
E(M_G^2-4E^2)\varphi_{ij}(P,{\bf q})= \hskip 3.2in
$$
\begin{equation}
i(\delta_{i i^\prime}-\frac{q_i
q_{i^\prime}}{{\bf q}^2})(\delta_{j
j^\prime}-\frac{q_jq_{j^\prime}}{{\bf q}^2})\int \frac{d^3k}{(2\pi)^3}
G_{i^\prime j^\prime k
l}(P,{\bf q,k})\varphi_{k l}(P,{\bf k}).
\end{equation}
where $E=\sqrt{q^2+m^2}$($m$ is the constituent mass of gluons, which will be
chosen as zero in the present calculations), and
$\varphi_{ij}(i,j=1,2,3)$ is the three dimensional wave function, which is related 
to $\chi_{ij}$ by the following equation
\begin{equation}
\varphi_{\mu \nu}(P,{\bf k})\equiv \int dk_0 \chi_{\mu \nu}(P,k).
\end{equation}
The kernel is divided into two parts: the short distance part
$G^{(s)}_{ijkl}$ and the long distance part $G^{(l)}_{ijkl}$.
$G^{(s)}_{ijkl}$ is obtained approximately by calculating the three 
lowest order diagrams shown in fig.1.\\
Besides equations (4), we also have
\begin{equation}
P=p_3+p_4,{~~~~}2k=p_3-p_4.
\end{equation}
Calculating the diagrams $a,b$ and $c$ shown in fig.1, $G^{(s)}_{\mu \nu \rho \sigma}(P,q,k)$
can be expressed explicitly as
$$
\displaystyle G^{(s)}_{\mu \nu \rho \sigma}(P,q,k)=
3i(4\pi \alpha_s)\left\{2C_{\mu \rho
\tau}(p_1,p_3)C_{\nu \sigma \tau^\prime}(p_2,p_4)
\displaystyle\left[ \frac{g^{\tau 0} g^{\tau^\prime 0}}{{\bf l}^2}
+\frac{g^{\tau i }g^{\tau^\prime j}}{l^2}(\delta_{ij}
-\frac{l_i l_j}{{\bf l}^2}) \right]\right.\\$$
\begin{equation}
\displaystyle  - \left. (2g_{\mu \nu}g_{\rho\sigma}
\displaystyle - g_{\mu \rho}g_{\nu \sigma}-g_{\mu \sigma}g_{\nu
\rho})\frac{}{}\right\},\hskip 1.0in
\end{equation}
where $l(=q-k)$ is the momentum exchanged between the two constituent gluons, 
and
\begin{equation}
C_{\mu \rho \tau}(p_1,p_3)=(p_1-2p_3)_\mu g_{\rho \tau}+(p_1+p_3)_\tau
g_{\mu \rho}+(p_3-2p_1)_\rho g_{\mu \tau},
\end{equation}
\begin{equation}
C_{\nu \sigma \tau^\prime}(p_2,p_4)=(p_2-2p_4)_\nu g_{\sigma
\tau^\prime}+(p_2+p_4)_{\tau^\prime}
g_{\nu \sigma}+(p_4-2p_2)_\sigma g_{\nu \tau^\prime}.
\end{equation}
The factor 3 on the right-hand side of equation (8) is the color factor which
is $\frac{4}{3}$ in the case of $q\bar q$ bound state.  Diagram $a$ and
diagram $b$ make the same contribution to physical states, so there is
a factor 2 in equation (8). The strong coupling constant $\alpha_s$ is chosen
as a running one,
\begin{equation}
\alpha_s=\frac{12\pi}{27}\frac{1}{ln(a+\frac{l^2}{\Lambda^2_{QCD}})},
\end{equation}
where $a$ is a parameter introduced to avoid the infrared divergence.  We
take $a$=4.0, which implies that the largest value of the running coupling
constant is $1.0$, and we choose  $\Lambda_{QCD}=200 MeV$.

As for the long distance part $G^{(l)}_{ijkl}$, we only choose in the
three-gluon-vertex (9,10) 
the terms  
containing tensor $g_{\mu \rho}g_{\nu \sigma}$ as the spin dependence of the
confining part, because such terms have nothing to do with spin effect.
Therefore we assume
\begin{equation}
G^{(l)}_{\mu \nu \rho \sigma}=2i(p_1+p_3)\cdot
(p_2+p_4)g_{\mu \rho}g_{\nu \sigma}G(l),
\end{equation}
where $G(l)$ is spatial dependence of confining part of the kernel. We choose
\begin{equation}
\displaystyle G(l)=\frac{8\pi\lambda}{l^4},
\end{equation}
which corresponds to a linearly growing potential. $\lambda$ is the string
tension, which in quark potential model takes the value $0.18(GeV)^2$.
However, here we chose $\lambda=0.36(GeV)^2$. We expect that
confining force between two gluons is one times stronger than that between two
quarks because  each gluon acts like a $q\bar{q}$ pair[4].
Of course, such a potential is very singular at the point $l=0$. Some form of
regularization is necessary. Our method is given in ref.\cite{s5}.
For $0^{++}$ glueball, the B.S. wave function can be expressed as
\begin{equation}
\varphi_{ij}(P,{\bf q})=f_s({\bf q})(\delta_{ij}-\displaystyle\frac{q_iq_j}{{\bf q}^2}).
\end{equation}
With the kernel and the form of the wave functions chosen as above, the
equation can be solved numerically. The glueball spectrum obtained are in
good agreement with that of lattice calculation. At the same time the wave
functions (14) and (15) are also obtained. Such wave functions should be 
normalized before being used in the calculation of the decays.
If we neglect the $P_0$ dependence of the kernel, then
the three dimensional wave function $\varphi_{ij}(P,{\bf k})$ is
subjected to the normalization condition
\begin{equation}
\displaystyle\int\frac{d^3{\bf q}}{(2\pi)^3}E\varphi_{ij}({\bf q})
\varphi_{kl}({\bf q})(\delta_{ik}
-\displaystyle\frac{q_i q_k}{{\bf q}^2})(\delta_{jl}-\frac{q_j q_l}{{\bf q}^2})
=(2\pi)^2
\end{equation}
\par
3.{\it The Width of  $0^{++}$ Glueball to Two Liglt Mesons} 
\par
i. {\it The decay to two pions}{~~~}
In the framework of perturbative QCD, the decays of the glueball
to two pions can be expressed by the diagram shown in fig.2.
The amplitude is expressed as
\begin{equation}
M(G(0^{++})\rightarrow\pi^+\pi^-)=
-\displaystyle\frac{4\pi \bar{\alpha}_s f_\pi^2}{9}
\int^1_{-1}d\xi_1 d\xi_2 \varphi_\pi(\xi_1)\varphi_\pi(\xi_2)(2k_{1i}
k_{2j}+k_1\cdot k_2\delta_{ij})\chi_{ij}(q).
\end{equation}
where $\bar{\alpha}_s$ is effective coupling constant which will be
discussed below, $k_1$, $k_2$ are the momentum of pions. 
$\varphi_\pi(\xi)$ is the pion wave function or distribution amplitude.
$\chi_{ij}$ is the four dimensional glueball B.S. wave function defined in
section 1., which
can be replaced by the three dimensional wave function $\varphi_{ij}$ through
\begin{equation}
\chi_{ij}=
\displaystyle\frac{|{\bf q}|(M_G^2-4{\bf q}^2)}{2\pi i p_1^2 p_2^2}\varphi_{ij}
\end{equation}
$\varphi_{ij}$ is obtained by solving the B.S. equation.

\par
The mass of pions can be neglected, and thus we have
\begin{equation}
\displaystyle k_1\cdot k_2=\frac{M_G^2}{2}.
\end{equation}
\begin{equation}
\displaystyle |{\bf q}|=\frac{|\xi_1+\xi_2|}{4}M_G.
\end{equation}
With these relations, equation (16) is reduced to
$$
M(G(0^{++})\rightarrow\pi^{+}\pi^{-})=
\frac{4i\bar{\alpha}_sM_G f_\pi^2}{9}\int_{-1}^{1}d\xi_1 d\xi_2
\varphi_{\pi}(\xi_1)\varphi_{\pi}(\xi_2) \hskip 1.4in$$
\begin{equation}
\hskip 0.6in
\cdot\frac{|\xi_1+\xi_2|[1-\frac{1}{2}
(\xi_1+\xi_2)^2]}{(1-\xi^2)(1-\xi^2)}\\
\displaystyle\sum_i\varphi_{ii}(|{\bf q}|)
\end{equation}
There are several forms of the pion wave function. Here we will use the form
given by Chernyak by QCD sum rule\cite{s9}.
\begin{equation}
\displaystyle\varphi_\pi(\xi)=\frac{15}{4}\xi^2(1-\xi^2).
\end{equation}

In equations (16) and (20), $\bar{\alpha}_s$ is the
effective coupling constant. In fact, we should have taken
$\sqrt{\alpha_s \cdot\alpha_s}$ at the points corresponding to the gloun virtualities:
 $\displaystyle\alpha_s(p_1^2)=\alpha_s(\frac{1-\xi_1}{2}\frac{1-\xi_2}{2}M_G^2)$ and
 $\displaystyle\alpha_s(p_2^2)=\alpha_s(\frac{1+\xi_1}{2}\frac{1+\xi_2}{2}M_G^2).$
However, for simplicity, we replace $p^2_{1,2}$ by their character values
$\bar{p}_{1,2}^2$, and then extract $\alpha_s$ out of the integrals over
$\xi_1$ and $\xi_2$. $\bar{p}^2_{1,2}$ can be determined by the forms of pion
wave functions. Therefore, $\bar{\alpha_s}$ in formula (16) is chosen as
$\displaystyle\frac{\alpha_s(\bar{p_1}^2)\alpha_s(\bar{p_2}^2)}{\alpha_s(M_G^2/4)}$.
For Chernyak's form, the pion wave function have maximum at $\bar{\xi}$=0.7,
and
$\displaystyle\bar{p}_{1,2}=\frac{1\mp\bar{\xi_1}}{2}\frac{1\mp\bar{\xi_2}}{2}M_G^2$.
Therefore, $\bar{\alpha}_s=0.69$ for $0^{++}$ glueball. 

\par Substituting expression (14) into formula (20), we get
\begin{equation}
 M(G(0^{++})\rightarrow \pi^+\pi^-)=
\displaystyle\frac{8i\bar{\alpha}_s f^2_\pi M_GI}{9}
\end{equation}
with
$$
I=\displaystyle \int_{-1}^{1}d\xi_1 d\xi_2
\varphi_{\pi}(\xi_1)\varphi_{\pi}(\xi_2)
\displaystyle\frac{|\xi_1+\xi_2|[1-\frac{1}{2}
(\xi_1+\xi_2)^2]}{(1-\xi^2)(1-\xi^2)}
f_s(|{\bf q}|)
$$
 Finally, after making the phase space integration, we obtain
\begin{equation}
\Gamma(G(0^{++})\rightarrow\pi^+\pi^-)
=\frac{4\bar{\alpha}_s^2 M_G f_\pi^4 I^2}{81\pi }
\end{equation}
The numerical results is shown in table 1.
\par
ii.{\it The decays to two $K$'s and two $\eta$'s}{~~~}
To estimate the glueball decays to two $k$ and two $\eta$ mesons, we
construct the following chiral Lagrangian in the lowest order
\begin{equation}
L_{eff} = g \phi tr(\partial_\mu\Sigma^{+}\partial^\mu\Sigma)
\end{equation}
where $\phi$ is the field of the scalar meson, and 
$\displaystyle\Sigma={\Large e^{\frac{2i\stackrel{\sim}{\pi}}{f_\pi}}}$, 
$$
\stackrel{\sim}{\pi}=\displaystyle\left(\begin{array}{ccc}
\displaystyle\frac{\pi^0}{\sqrt{2}}+\frac{\eta}{\sqrt{6}}&\pi^+ &k^+ \\
\pi^- & -\displaystyle\frac{\pi^0}{\sqrt{2}}+\frac{\eta}{\sqrt{6}}& k^0\\
k^- &\bar{k}^0 & -\displaystyle\frac{2\eta}{\sqrt{6}}\\
\end{array}\right)
$$
$g$ is a coupling constant which can be determined from  the result of
two pions decay obtained above. Neglecting the factor of the phase space,
we have
$$\Gamma(\pi^+\pi^-)=\Gamma(k^+ k^-)=
\Gamma(k^0\bar{k}^0)=\frac{1}{2}\Gamma(\pi^0\pi^0)
=\frac{1}{2}\Gamma(\eta\eta)$$
Taking the factor of the phase space into consideration, We give the
numerical results in table.1. 
\par
4.{\it The decay of glueball to $\eta$, $\eta^\prime$ form QCD anomaly}~~~
As mentioned in the introduction, $\eta$, $\eta^\prime$ final state cannot be 
produced enough only from naive quark model, there should be another
mechanism to produce them, namely the QCD anomaly (see fig.3)
\begin{equation}
\partial_\mu j_5^\mu=\frac{3\alpha_s}{4\pi}G_{\mu\nu}G^{\mu\nu}+
2i\sum\limits_{q=uds}m_q\bar{q}\gamma^5q
\end{equation}
where $j_5^\mu$ is the singlet axial current [10]. Indeed it is the anomalous
contribution to the divergence of the axial current that makes the $SU(3)$
singlet state much heavier than the Octet pseudo-Goldstone state and
distinguish $\eta$ from $\eta^\prime$. Obviously, keeping 
$SU(3)_L\otimes SU(3)_R$
flavor symmetry unbroken, $\eta$ cannot be produced from two gluons.
Actually, physical $\eta$, $\eta^\prime$ states are the mixing states of
flavor eigenstate $\eta_1$ and $\eta_8$
\begin{equation}
\eta=cos\theta_\eta \eta_8 - sin\theta_\eta \eta_1,
\end{equation}
\begin{equation}
\eta^\prime=sin\theta_\eta \eta_8 + cos\theta_\eta \eta_1,
\end{equation}
where the mixing angle $\theta_\eta$ is estimated in the range of $-10^\circ$ to
$-20^\circ$, so $\eta^\prime$ is dominantly $\eta_1$.
The effective coupling of $\eta_1 gg$ can be written in the form
\begin{equation}
H(q_1^2,q_2^2,q_{\eta_1}^2)\delta^{ab}\epsilon_{\mu\nu\alpha\beta}q^\mu_1
q^\nu_2\epsilon_1^\alpha\epsilon_2^\beta/cos\theta_\eta
\end{equation}
where $q_1$, $q_2$ and $\epsilon_1$, $\epsilon_2$ are the 4-momenta and
polarizations of two gluons  and  $a$, $b$ are color indices. $H$ is a form
factor that is a general function of momenta, $q_1^2$, $q_2^2$ and $q_\eta^2$.
For simplicity,  we set $H$ as a constant , then, from the
experimental data $\Gamma(J/Psi\rightarrow
\gamma\eta')/\Gamma(J/\Psi\rightarrow e^+e^-)=7\times 10^{-2}$,
one may obtain  $H\sim1.8$ GeV$^{-1}$ \cite{s11}.
Corresponding to fig.3, the amplitude dominated by on-shell gluons can be
written as
$$
M(G\rightarrow\eta\eta^\prime)
=3\sqrt{8}cos\theta_\eta sin\theta_\eta \int\frac{d^4 q}{(2\pi)^4}
\frac{d^3q_1}{(2\pi)^3 2|{\bf p}_1|}\frac{d^3q_2}{(2\pi)^3 2|{\bf p}_2|}
\frac{d^3q_3}{(2\pi)^3 2|{\bf p}_3|}\frac{d^3q_4}{(2\pi)^3 2|{\bf p}_4|}
\hskip 1.2in$$
$$
\cdot(2\pi)^4\delta^4(\frac{1}{2}P+q-q_1-q_2)(2\pi)^4\delta^4(k_1-q_1-q_3)
(2\pi)^4\delta^4(k_2-q_2-q_4)$$
\begin{equation}
\cdot\chi_{\mu\nu}(P,q)C^{\rho\sigma\mu}(-q_1,q_2)C^{\alpha\beta\nu}(-q_3,q_4)H^2
\epsilon_{\rho\alpha\gamma\delta}\epsilon_{\sigma\beta\lambda\tau}q_1^\gamma
q_2^\lambda q_3^\delta q_4^\tau \hskip 0.7in
\end{equation}
where $C^{\rho\sigma\mu}(-q_1,q_2)$ and $C^{\alpha\beta\nu}(-q_3,q_4)$ are the
three-gluon-vertex defined as eq.(9), and $3\sqrt{8}$ is a color factor.
With the the glueball B.S. wave function obtained in section 2, the
calculation of the amplitude is straight forward. The corresponding width is
list in table 1.
\par
Because the sum of $m_\eta$ and $m_{\eta^\prime}$ is quite near the mass of
glueball, $M$, we expect the above result from QCD anomaly will
depend on $M$.
On the other hand, the glueball mass is not exactly known either from
experiment or from theory. Therefore, we show the dependence of
$\Gamma_{G\rightarrow\eta\eta^\prime}$ on glueball mass $M$ in fig.4.
The value of $\Gamma(\eta\eta_\prime)$ in table 1 corresponds to $M=1.6$GeV.
\par
5.{\it Discussion and Conclusions}\\
In this paper, we study the decays of $0^{++}$ glueball to two light
mesons by using the constituent gluon model. The glueball wave function is 
obtained by solving B-S equation, the pion is light so that it can
described by the light cone wave function. SU(3) chiral symmetry is used
to predict decay widths $\Gamma(\bar K K)$ and $\Gamma(\eta\eta)$. 
we also study the  $\eta$ $\eta^\prime$ production  from QCD anomaly, 
which almost is zero from the naive quark model prediction.    
We are more interested in the ratios of the different decay channel, since 
the ratios is less sensitive to the glueball wave function we use. 
From table 1, we get
$$\Gamma(\eta\eta^\prime)/\Gamma(\eta\eta)=0.82 ~~~~~
\Gamma(\pi^0\pi^0)/\Gamma(\eta\eta)=1.5$$
Such ratios have values in  experiment for $f_0(1500)$. The Crystal Barrel 
Collab.\cite{s12} gives
\begin{eqnarray}
\Gamma(\eta\eta^\prime)/\Gamma(\eta\eta)=0.29\pm0.1\\
\Gamma(\pi^0\pi^0)/\Gamma(\eta\eta)=1.45\pm0.61 ( 3\pi^0 channel)\\
\Gamma(\pi^0\pi^0)/\Gamma(\eta\eta)=2.12\pm0.81 (Coupled-channel),
\end{eqnarray}
GAM group gave a rather large ratio \cite{Bin}
\begin{equation}
\Gamma(\eta\eta^\prime)/\Gamma(\eta\eta)=2.7\pm0.8
\end{equation}

We see that our results agree with the experiment data of
$f_0(1500)$.

\newpage

Figuer Caption\\
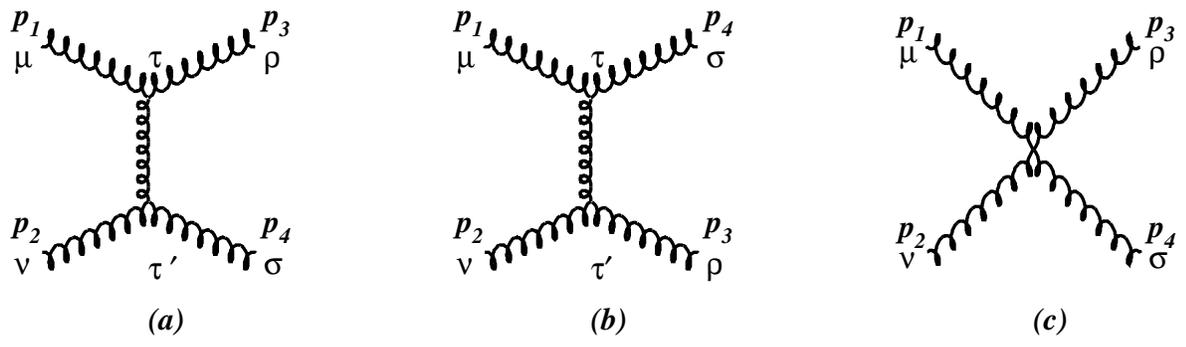
Fig.1: The diagrams contributing to the short distance part of the B.S.
kernel.\\
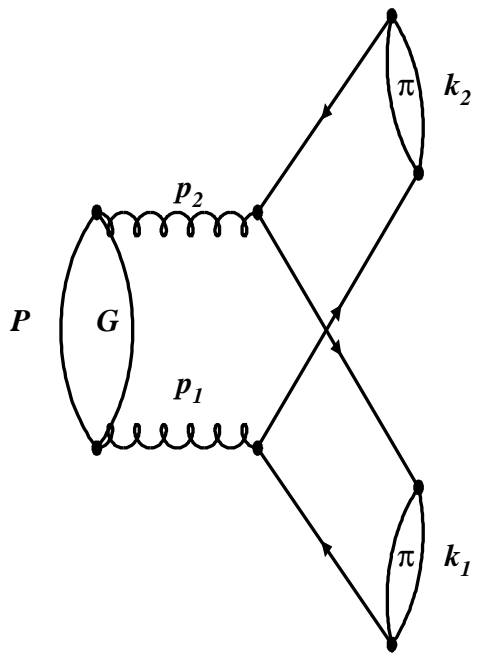
Fig.2: The decays of glueball to two pions in purterbative framework.\\
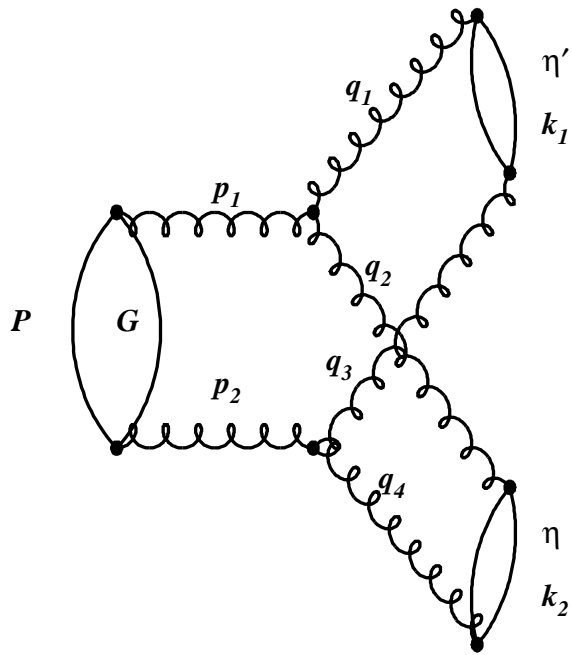
Fig.3: The decays of glueball to $\eta\eta^\prime$ from QCD anomaly\\
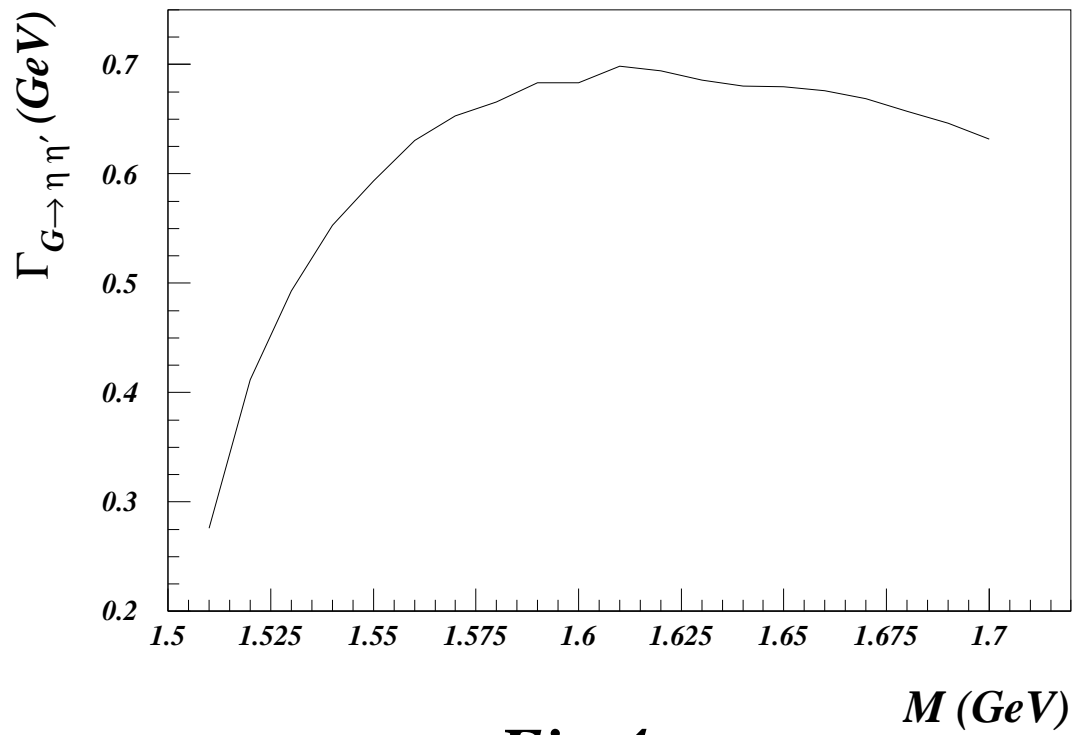
Fig.4: The dependence of $\Gamma(\eta\eta^\prime)$ on glueball mass. 
\newpage
Table Caption\\

Table 1: The partial width of the  glueball to two light mesons (MeV). 

\newpage
\begin{thebibliography}{9}
\bibitem{s1}UKQCD Colloboration, G. Bali, K. Schilling, A. Hulsebos, 
A.C. Irving,C. Michael and P. Stephenson, Phys. Lett. B309(1993)378;

\bibitem{s2}H. Chen,J. Sexton, A. Vaccarino and D. Weingarten, 
Nucl. Phys. B(Proc. Suppl.) 34(1994)357;
\bibitem{s3}C. Amsler and F.E. Close, Phys. Lett. B353(1995)385;
Phys. Rev. D53(19996)295;
\bibitem{s4}J.M. Cornwall and A. Soni, Phys. Lett. B120(1983)431;
\bibitem{s5}J.Y. Cui, J.M. Wu and H.Y. Jin, HEP-PH/9711379(submitted to
Phys. Lett. B)
\bibitem{s6}Crystal Barral Collaboration, Phys. Lett. B355(1995)425;

\bibitem{s7}D.V. Bugg at al., Phys. Lett. B353(1995)378;
\bibitem{s8}V.V. Anisovich et al., Phys. Lett. B323(1994)233;
\bibitem{s9}V.L. Chernyak and A.R. Zhitinisky, Phys. Rep. 112 (1984)175;
\bibitem{s10}C. Itzyson and J.B. Zuber, Quantum Field Theory (McGraw-Hill,
New York, 1980);
\bibitem{s11}D. Atwood, A. Soni, Phys. Lett. B405(1997)150;
\bibitem{s12}C. Amsler at al., Phys. Lett. B353(1995)571; 
\bibitem{Bin}Particle Data Book, Phys.Rev{\bf D 54}. 
\end{thebibliography}
\end{document}